\documentclass[aps,prl,showpacs,twocolumn,groupedaddress]{revtex4-1}

\usepackage{longtable}
\usepackage{graphicx}
\usepackage{dcolumn}
\usepackage{bm}
\usepackage{amsmath,amssymb}
\usepackage{longtable}

\newcommand{\PDG}{Beringer:1900zz}
\newcommand{\re}{\text{Re }}
\newcommand{\im}{\text{Im }}

\begin{document}


\title{Structure of Near-Threshold $s$-Wave Resonances}


\author{Tetsuo~Hyodo}
\email[]{hyodo@yukawa.kyoto-u.ac.jp}
\affiliation{Department of Physics, Tokyo Institute of Technology,
Tokyo 152-8551, Japan.}
\altaffiliation{Present address: Yukawa Institute for Theoretical Physics, Kyoto University, Kyoto 606-8502, Japan}


\date{\today}

\begin{abstract}
We study the structure of two-body $s$-wave bound states as well as resonances in the threshold energy region. We focus on the single-channel scattering where the scattering length and the effective range are given by real numbers. It is shown that, in the energy region where the effective range expansion is valid, the properties of resonances are constrained only by the position of the pole. We find that the compositeness defined through the analytic continuation of the field renormalization constant is purely imaginary and normalized for resonances. We discuss the interpretation of this quantity by examining the structure of the hadron resonance $\Lambda_{c}(2595)$ in the $\pi\Sigma_{c}$ scattering. We show that the $\Lambda_{c}(2595)$ resonance requires an unnaturally large effective range and hence it is not likely a $\pi\Sigma_{c}$ molecule.
\end{abstract}

\pacs{14.20.Gk,03.65.Nk,14.20.Lq}



\maketitle


The appearance of resonances is a common phenomenon in various fields in physics. Several types of unstable particles are generated by different mechanisms in particle, nuclear, and condensed matter physics~\cite{Kukulin}. In the strong interaction sector governed by quantum chromodynamics, hundreds of hadrons have been experimentally observed, most of which are unstable against the strong decays~\cite{\PDG}. It is therefore important to study the structure of hadron resonances, in order to understand the nonperturbative dynamics of the strong interaction. 

Although the typical behavior of resonance phenomena is explained in the textbooks of quantum mechanics, the theoretical foundation of the description of resonances has not been well established~\cite{Berggren:1968zz}. For instance, the field renormalization constant $Z$ has been used to study the structure of two-body bound states~\cite{Weinberg:1965zz}. The quantity $1-Z$ represents the compositeness of the bound state, which is well defined and normalized for stable bound states. Generalization of the compositeness to resonances has been formulated, for instance, by using the integration of the spectral density~\cite{Baru:2003qq,Hanhart:2011jz}. Another approach is to define the compositeness as the analytic continuation of the field renormalization constant of the resonance pole~\cite{Hyodo:2011qc,Aceti:2012dd}. This approach is a straightforward generalization of the bound state case, which is free from the background (nonresonant) contribution of the scattering. On the other hand, it provides a complex and unnormalized number for compositeness whose interpretation is not clear.

In this Letter, we consider this problem by employing the effective range expansion of an $s$-wave scattering, which is a model-independent expression of the amplitude as far as the small momentum region is concerned. Because the effective range expansion specifies the amplitude by two threshold quantities, the properties of near-threshold resonances are highly constrained. This enables us to extract the general feature of the near-threshold resonances.


Let us consider a single-channel scattering in the effective range expansion. Truncating the expansion of $k\cot\delta$ up to $k^{2}$, we write the scattering amplitude with the momentum $k$ as
\begin{align}
    f(k)
    = & \left(\frac{1}{a}-ki+\frac{r_{e}}{2}k^{2}\right)^{-1} ,
    \label{eq:amplitude}
\end{align}
where $a$ is the scattering length and $r_{e}$ is the effective range~\footnote{We adopt the convention of the scattering length as $f(k)\to a$ when $k\to 0$.}. The truncation should be valid in the small $k$ region, namely, the near-threshold kinematics. For a given set of ($a$, $r_{e}$), the amplitude~\eqref{eq:amplitude} has two poles at
\begin{align}
    k^{\pm}
    = & \frac{i}{r_{e}}\pm \frac{1}{r_{e}}\sqrt{-\frac{2r_{e}}{a}-1}
    \label{eq:solution} .
\end{align}
We call $k^{-}$ the ``primary pole'', which is closer to the physical scattering axis ($\im k=0$, $\re k\geq 0$). The other pole $k^{+}$ is called the conjugate pole. In the following, we consider the lowest energy threshold without any open channels below, so that $a$ and $r_{e}$ are real~\footnote{In fact, all the formulas can be applied to complex $(a, r_{e})$, but the interpretation of the obtained quantities is only possible for the real-valued threshold parameters. In addition, the inclusion of higher energy coupled channels is in principle straightforward, as described in Refs.~\cite{Baru:2003qq,Hanhart:2011jz,Hyodo:2011qc}.}. The positions and properties of these poles are classified in Ref.~\cite{Ikeda:2011dx}. As is well known, for a positive effective range  $r_{e}>0$, the primary pole represents a bound (virtual) state for the negative (positive) scattering length. The allowed region for the bound state is constrained by the condition $-2r_{e}<a$ because of the causality. Although Eq.~\eqref{eq:amplitude} always has two poles, only those that appear in the small $k$ region are physically relevant.

For a negative effective range $r_{e}<0$, the primary pole can represent a resonance state. Naively, a simple attractive potential does not generate an $s$-wave resonance, because there is no centrifugal barrier. This is understood because the simple attraction provides a positive effective range. A negative effective range can be realized, for instance, by energy dependent interactions, by nonlocal interactions, and by Feshbach resonances~\cite{Phillips:1997xu}. 

\begin{figure}[tbp]
    \centering
    \includegraphics[width=8cm,clip]{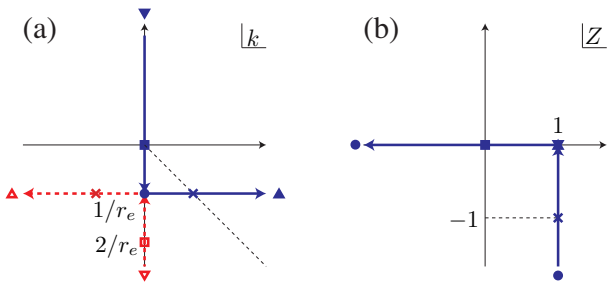}
    \caption{\label{fig:Schematic}
    (color online). Trajectories of the pole positions $k^{-}$ and $k^{+}$ (a), and the field renormalization constant $Z$ (b) increasing the inverse scattering length $1/a$ for a fixed negative effective range $r_{e}<0$. Inverted triangles, squares, circles, crosses, and triangles correspond to $1/a=-\infty$, $0$, $-2/r_{e}$, $-1/r_{e}$, and $+\infty$, respectively. Solid (dashed) line with filled (empty) symbols stands for $k^{-}$ ($k^{+}$).}
\end{figure}

With a fixed $r_{e}<0$, we plot the trajectories of the poles~\eqref{eq:solution} in the complex $k$ plane by varying the inverse scattering length $1/a$ from $-\infty$ to $+\infty$ in Fig.~\ref{fig:Schematic}(a). The primary pole moves from the bound state region to the virtual state region, and merges with the conjugate pole at the double root $k^{-}=k^{+}=i/r_{e}$. The pole then moves off the imaginary $k$ axis while acquiring a real part, and eventually turns into a resonance. We note that the double root should lie in the negative region of the imaginary $k$ axis, in order to have a resonance ($k^{-}$ in the fourth quadrant). The properties of the poles are summarized in Table~\ref{tbl:classification}.

\begin{table*}[bt]
\caption{Classification of the properties of the poles in the effective range expansion for $r_{e}<0$.}
\begin{center}
\begin{ruledtabular}
\begin{tabular}{lcccc}
Inverse scattering length & $1/a<0$ & $0< 1/a<-2/r_{e}$ & $-2/r_{e}< 1/a<-1/r_{e}$ & $-1/r_{e}< 1/a$ \\
\hline
Primary pole & Bound state & Virtual state & Virtual state with width & Resonance  \\
Conjugate pole & Virtual state & Virtual state & Anti-virtual state with width & Anti-resonance \\
Compositeness & $0<X<1$ & $1< X$ & $1<\bar{X}$ & $0<\bar{X}<1$ \\
\end{tabular}
\end{ruledtabular}
\end{center}
\label{tbl:classification}
\end{table*}%

The scattering length and the effective range can be expressed by the pole positions as
\begin{align}
    a
    = &\frac{k^{+}+k^{-}}{ik^{+}k^{-}},
    \quad
    r_{e}
    =  \frac{2i}{k^{+}+k^{-}} \label{eq:are} .
\end{align}
In the present case, because $k^{+}+k^{-}$ ($k^{+}k^{-}$) is purely imaginary (real), both $r_{e}$ and $a$ are real numbers. The scattering amplitude is then written as $f(k)=(k^{+}+k^{-})/[i(k-k^{+})(k-k^{-})]$ so the residue of the pole is obtained as
\begin{align}
    \lim_{k\to k^{\pm}}(k-k^{\pm})f(k)
    = & \frac{k^{+}+k^{-}}{i(k^{\pm}-k^{\mp})}
    \label{eq:residue} .
\end{align}
Again, this is a \textit{real} number. For the bound and virtual states, the residue of the primary pole $k^{-}$ is determined by the position of the conjugate pole $k^{+}$ and vise versa. In the case of the resonances, $k^{+}=-(k^{-})^{*}$, so the residue is solely determined by the position of the pole. Note that in general the residue of the resonance pole is a complex number, which is independent of the pole position. Equation~\eqref{eq:residue} suggests that the properties of the near-threshold resonances are constrained through the threshold quantities.


Next we turn to the compositeness. For a weakly bound state, the scattering length and the effective range are related to the field renormalization constant as~\cite{Weinberg:1965zz} $a= -2(1-Z)R/(2-Z)$ and $r_{e}=-ZR/(1-Z)$ with $R=(2\mu B)^{-1/2}$, the reduced mass $\mu$, and the binding energy $B$. The field renormalization constant $Z$ is defined as the overlap of the physical bound state with the elementary contribution other than the scattering state. By eliminating $R$, $Z$ is given by 
\begin{align}
    Z
    = & 1-\sqrt{
    1-\frac{1}{1+a/(2r_{e})}}
    =
    \frac{2k^{-}}{k^{-}-k^{+}}
    \label{eq:Zare} ,
\end{align}
where we choose the sign of the square root so that the expression matches with the normalization $0< Z< 1$ for the bound states. The quantity $X\equiv 1-Z$ is called compositeness, which measures the two-body molecule component in the bound state. 

Now we consider the $1/a$ dependence of $Z$ with a fixed $r_{e}<0$ [Fig.~\ref{fig:Schematic}(b)]. It is instructive to consider $X=1-Z$, instead of $Z$ itself. The compositeness $X$ is real and positive for $1/a<-2/r_{e}$. In addition, for the bound states (negative $1/a<0$), the compositeness is always normalized as $0<X<1$. We obtain a pure composite state $X=1$ in the unitary limit $1/a=0$. Beyond the unitary limit, the bound state turns into a virtual state and the compositeness exceeds unity. At the double root $1/a=-2/r_{e}$, $X$ diverges and for $1/a>-2/r_{e}$, $X$ becomes purely imaginary. Interestingly, however, its magnitude is normalized within $0<|X|<1$ for the resonance case $-1/r_{e}< 1/a$. Here we define a new quantity
\begin{align*}
    \bar{X}\equiv
    & \sqrt{
    -1+\frac{1}{1+a/(2r_{e})}} ,
\end{align*}
which is real and positive for $1/a>-2/r_{e}$, and properly normalized for resonances ($0<\bar{X}<1$). The normalization is given at $\bar{X}=1$ at $ k^{-}=-1/r_{e}+i/r_{e}$ with $1/a=-1/r_{e}$. The corresponding eigenenergy is 
\begin{align}
    E^{-}
    =
    \frac{(k^{-})^{2}}{2\mu}
    = -\frac{i}{\mu r_{e}^{2}} 
    \label{eq:thwidth} .
\end{align}
This is a special state whose mass is located at the two-body threshold and the width is determined solely by the effective range; the width is small (large) for a large (small) $|r_{e}|$. $\bar{X}=0$ is realized when $\re k^{-}\to +\infty$ ($1/a\to +\infty$). The effective range expansion, however, breaks down before we take this limit. 


We can summarize the situation as follows. In the effective range approximation, we have two parameters, $(a,r_{e})$. For a given position of the bound state, the pair $(a,r_{e})$ is related to the position of the bound state pole and its residue~\eqref{eq:residue}. As shown in Ref.~\cite{Weinberg:1965zz}, the compositeness $1-Z$ can be calculated by these quantities. Furthermore, $(a,r_{e})$ also determines the position of the conjugate pole~\eqref{eq:solution}. As discussed in Ref.~\cite{Baru:2003qq}, the distance between $k^{-}$ and $k^{+}$ is related to the compositeness, in accord with the ``pole counting'' argument~\cite{Morgan:1990ct}. If two poles appear close to each other [$k^{+}\sim k^{-}$ in Eq.~\eqref{eq:Zare}], $Z\sim 1$ and the bound state is considered to be an elementary contribution other than the two-body state. In this case, $k^{+}$ can be regarded as the shadow pole~\cite{Eden:1964zz}.

For the resonances, the field renormalization constant $Z$ is calculated by the position of the pole and the residue~\cite{Hyodo:2011qc,Aceti:2012dd}. In general, the residue of the pole is independent of its position. However, the residue of a near-threshold resonance is related to its pole position through the effective range parameters, so $Z$ is determined only by the pole position. This is a universal feature of the near-threshold resonances.


To examine the validity of the effective range expansion, we consider a single-channel model for the $s$-wave pion-baryon scattering based on chiral dynamics~\cite{Ikeda:2011dx,Hyodo:2011ur}. The scattering amplitude with the total energy $W$ is given by $T(W)=[V(W)^{-1}-G(W;d)]^{-1}$ where $G$ is the two-body loop function with the subtraction constant $d$ which specifies the finite part. The low-energy interaction $V(W)$ is model-independently determined by the chiral low energy theorem~\cite{Weinberg:1966kf} as
\begin{align}
    V(W)
    =& -\frac{C}{2f^{2}}(W-M)\frac{M+E(W)}{2M}
    \label{eq:interaction}
\end{align}
where $E(W)=(W^{2}-m^{2}+M^{2})/2W$, $f=92.4$ MeV is the pion decay constant. For the later application to $\Lambda_{c}(2595)$, we consider the $\pi\Sigma_{c}$ scattering and adopt the baryon mass $M=2453.54$ MeV and the pion mass $m=138.04$ MeV. The low energy theorem determines the group theoretical factor as $C=4$ for the isospin singlet channel, but we vary it to examine its dependence.

\begin{figure}[tbp]
    \centering
    \includegraphics[width=8cm,clip]{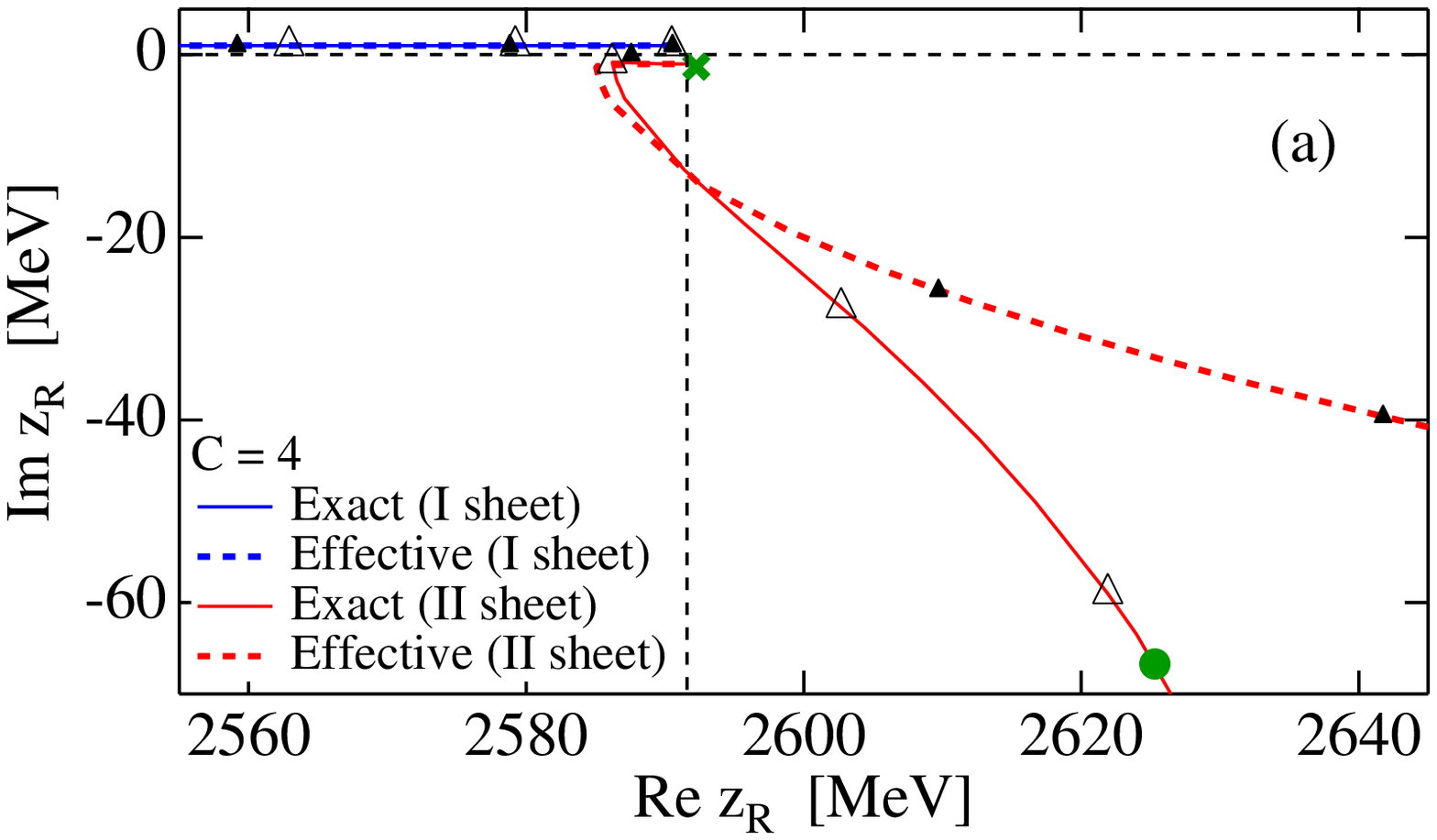}
    \includegraphics[width=8cm,clip]{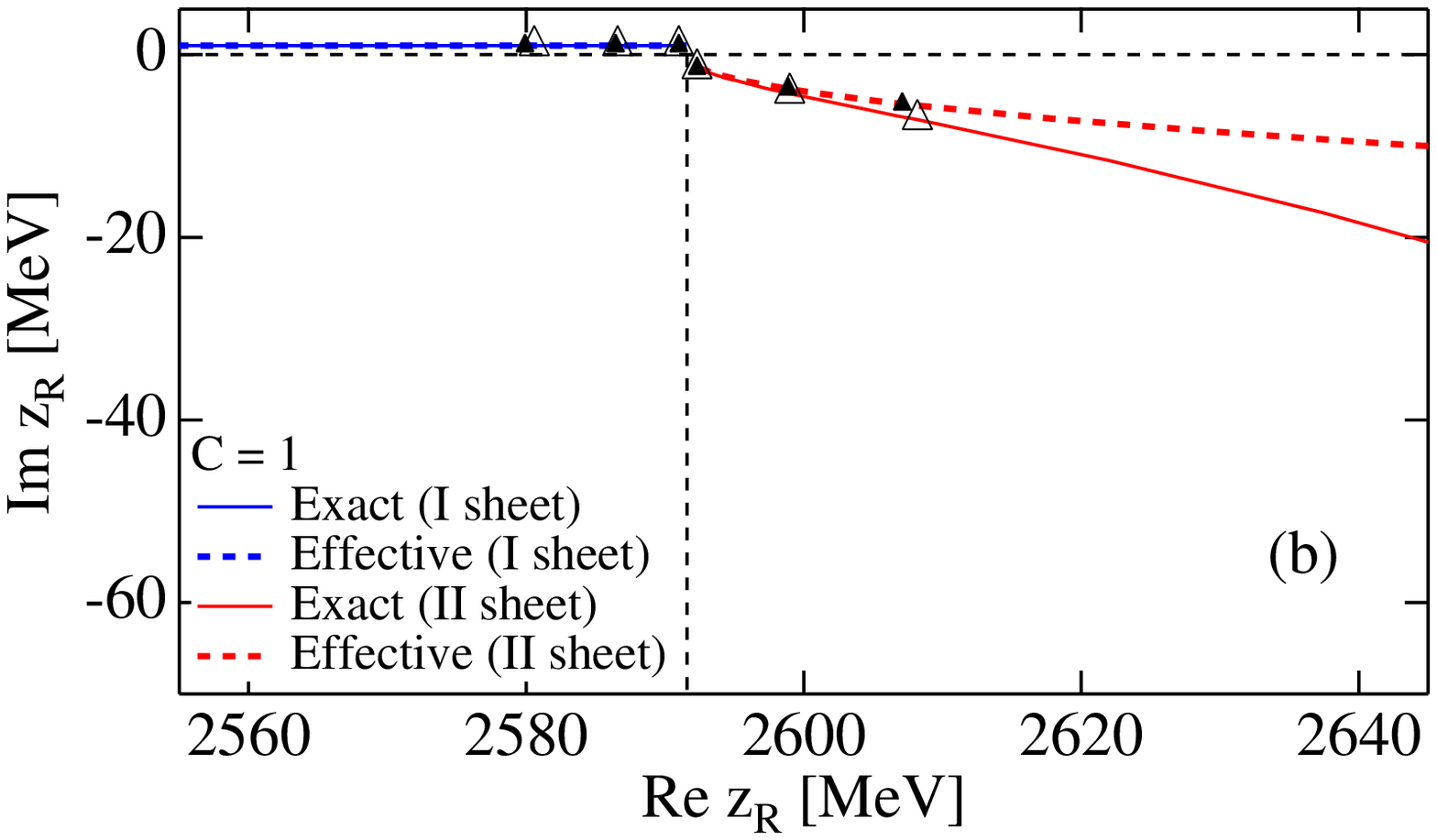}
    \caption{\label{fig:Pole}
    (color online). Trajectories of the pole position $z_{R}$ with the interaction strength $C=4$ (a) and $C=1$ (b). Solid (dashed) lines with empty (filled) triangles represent the exact pole positions [pole positions calculated from  ($a$, $r_{e}$)]. Symbols are plotted at $a=\pm 1$, $\pm 2$, and $\pm 10$ fm. Cross and circle in (a) represent the position of the physical $\Lambda_{c}(2595)$ and that of the natural renormalization scheme.}
\end{figure}

For a given interaction strength $C$ and the subtraction constant $d$, we calculate the pole position $z_{R}$ in the complex energy plane and ($a$, $r_{e}$) at the threshold. In Fig.~\ref{fig:Pole}, we plot by solid lines the trajectories of the pole position $z_{R}$ with respect to the variation of $d$. The interaction strength is fixed at $C=4$ (a) and $C=1$ (b).

As shown in Ref.~\cite{Ikeda:2011dx}, the effective range is stable against the change of the subtraction constant $d$, while the scattering length strongly depends on $d$. This is because the scattering length (effective range) is determined by the strength (derivative) of the inverse amplitude at the threshold. The change of the subtraction constant effectively modifies the strength of the inverse amplitude~\cite{Hyodo:2008xr}, while the energy dependence is not very much affected. In fact, the variation of the effective ranges is within $\pm 0.2$ fm in all cases in Fig.~\ref{fig:Pole}, while the scattering length changes its sign. Thus, the trajectories in Fig.~\ref{fig:Pole} effectively correspond to the scattering length dependence of the pole position with a fixed effective range. The central values are $r_{e}\sim -4.8$ fm and $\sim -18.3$ fm for $C=4$ and $C=1$, respectively. 

We show the pole trajectories calculated from ($a$, $r_{e}$) with Eq.~\eqref{eq:solution} by dashed lines in Fig.~\ref{fig:Pole}. The triangles are plotted at $a=\pm 1$, $\pm 2$, and $\pm 10$ fm. The different positions of the symbols, as well as the deviation of the dashed and solid lines indicate the validity of the effective range approximation. We thus confirm that the effective range expansion works well in the threshold energy region, especially when the magnitude of the effective range is large. Figure~\ref{fig:Pole}(a) shows that the expansion is also applicable to  the state whose excitation energy is smaller than its width, as far as the state is located close to the threshold.

It is worth noting that the residue of the pole in the amplitude with the normalization~\eqref{eq:interaction} is given by
\begin{align}
    g^{2}
    = & -\frac{2\pi (k^{+}+k^{-})^{2}}
    {\mu^{2}Mi(E^{-}-E^{+})} 
    [(E^{-}+M+m)k^{-}]
\end{align}
where $E^{\pm}=(k^{\pm})^{2}/(2\mu)$. The first factor is real, while the second one is complex. Nevertheless, the residue is uniquely determined by the pole position.


We now consider $\Lambda_{c}(2595)$, a negative parity excited state of the charmed baryon. In the second row of Table~\ref{tbl:L2595}, we show the central values of the mass and width of $\Lambda_{c}(2595)$ from the Particle Data Group~\cite{\PDG}. In view of Fig.~\ref{fig:Pole}, the pole position is well within the applicability of the effective range expansion, so we calculate the corresponding scattering length and the effective range by Eq.~\eqref{eq:are}. Using Eq.~\eqref{eq:Zare} we obtain the field renormalization constant $Z= 1-0.608 i$ and $\bar{X}=0.608$. To interpret this quantity, we examine the structure of $\Lambda_{c}(2595)$. 

\begin{table}[btp]
\caption{Comparison of the properties of the physical $\Lambda_{c}(2595)$ resonance and $\pi\Sigma_{c}$ threshold quantities with those of natural renormalization scheme. $E_{R}=\re z_{R}-M-m$ and $\Gamma_{R}=2\ \im z_{R}$.}
\begin{center}
\begin{ruledtabular}
\begin{tabular}{lcccc}
 & $E_{R}$ (MeV) & $\Gamma_{R}$ (MeV) & $a$ (fm) & $r_{e}$ (fm) \\
\hline
Physical & $0.67 $ & $2.59$ & 10.5 & $-19.5$  \\
Natural & $33.70$ & $133.31$ & 0.9 & $-4.6$  \\
\end{tabular}
\end{ruledtabular}
\end{center}
\label{tbl:L2595}
\end{table}%

The natural renormalization scheme was introduced in Ref.~\cite{Hyodo:2008xr} to exclude possible Castillejo-Dalitz-Dyson (CDD) pole contributions, which stem from the contributions other than the considered model space~\cite{Castillejo:1956ed}. Resonances generated in this scheme are regarded as hadronic molecule states. The subtraction constant is determined in this scheme as $d_{\rm natural} = -2.88$ at $\mu=630$ MeV, and we set the coupling strength $C=4$ to match with the low energy theorem. The pole positions are shown in Fig.~\ref{fig:Pole}(a) and Table~\ref{tbl:L2595}. Because the pole in the natural renormalization scheme is far from the physical one, we find that the substantial CDD pole contribution is required in the structure of $\Lambda_{c}(2595)$. 

This can also be confirmed by the scattering length and effective range. We show the threshold variables in the natural scheme in Table~\ref{tbl:L2595}. We find large deviation from the corresponding physical $\Lambda_{c}(2595)$, which again indicates substantial CDD pole contribution. In addition, the effective range $-19.5$ fm deduced from the physical pole position is an order of magnitude larger than the typical scale of the hadronic interaction of several fm. Thus, the effective range is at odds with the interpretation of a hadronic molecule. In view of these results, we conclude that the $\Lambda_{c}(2595)$ resonance is not dynamically generated by the chiral low energy interaction in the $\pi\Sigma_{c}$ channel. The origin of the CDD pole contribution may be a three-quark state, bound states of other channels such as $\pi\Sigma_{c}^{*}$, $DN$, $\pi\pi\Lambda_{c}$ and their mixtures.

In this analysis, we have assumed isospin symmetry for particle masses, and ignored the decay width of the $\Sigma_{c}$ baryon $\Gamma_{\Sigma_{c}}\sim 2$ MeV and the effect of the $\pi\Sigma_{c}^{*}$ channel which locates about 65 MeV above. The genuine $\pi\pi\Lambda_{c}$ three-body component is not included. In a more quantitative discussion, these effects should be taken into account. Nevertheless, the large deviation in Table~\ref{tbl:L2595} clearly disfavors the $\pi\Sigma_{c}$ molecule interpretation of the $\Lambda_{c}(2595)$ resonance.


Now we consider the interpretation of $\bar{X}$. For resonances ($1/a>-2/r_{e}$), we have $Z=1-i\bar{X}$. Since the field renormalization constant indicates how the bare propagator is modified from unity~\cite{Hyodo:2011qc}, we also expect that $\bar{X}$ represents the deviation from the elementary state $Z=1$. However, $\bar{X}=0$ is defined at $\re k^{-}\to +\infty$, which is far beyond the applicability of the effective range expansion. On the other hand, we have  discussed the special case where the real part of the resonance coincides with the threshold energy and the imaginary part is determined by the effective range. This resonance produces $\bar{X} = 1$ and it is well within the applicability of the effective range expansion. We therefore consider that $\bar{X}$ measures the deviation from this particular state.


In summary, we have studied the properties of the near-threshold resonances, analyzing the $\Lambda_{c}(2595)$ resonance as an example. We show that the effective range expansion is very useful to extract the properties of these resonances. As in the case of the bound states~\cite{Weinberg:1965zz}, the properties of the near-threshold resonances can be related to the \textit{observable} quantities. We find that the pole position of $\Lambda_{c}(2595)$ indicates a much larger effective range than the typical hadronic length scale, so the interpretation as a $\pi\Sigma_{c}$ molecule is not favored. Since the effective range expansion itself is a general expression for low energy scattering, the present framework can be applied to study the structure of any near-threshold resonances, such as $^{8}$Be in the $\alpha\alpha$ scattering. 

The author thanks Atsushi Hosaka, Daisuke Jido, Yusuke Nishida, Makoto Oka, and Toshitaka Uchino for fruitful discussions. This work is supported in part by Grant-in-Aid for Scientific Research from MEXT and JSPS (Grants No. 24105702 and No. 24740152).



\end{document}